\documentclass{ws-mpla}

\begin{document}

\markboth{A. M. Gavrilik and A. P. Rebesh}
{A $q$-oscillator with 'accidental' degeneracy of energy levels}


\title{\bf A $q$-OSCILLATOR WITH 'ACCIDENTAL' DEGENERACY OF ENERGY LEVELS}

\author{\footnotesize A. M. GAVRILIK}

\address{Bogolyubov Institute for Theoretical Physics, Kiev 03680, Ukraine\\
omgavr@bitp.kiev.ua
}

\author{A. P. REBESH}

\address{Bogolyubov Institute for Theoretical Physics, Kiev 03680, Ukraine\\
omgavr@bitp.kiev.ua
}

\maketitle

\pub{Received (Day Month Year)}{Revised (Day Month Year)}

\begin{abstract}
We study main features of the exotic case of $q$-deformed oscillators
(so-called Tamm-Dancoff cutoff oscillator) and find some special
properties:
(i) degeneracy of the energy levels $E_{n_1} = E_{n_1+1}$,
$n_1\ge 1$, at the {\em real value} $q=\sqrt{\frac{n_1}{n_1+2}}$ of
deformation parameter, as well as the occurrence of other degeneracies
$E_{n_1} = E_{n_1+k},\ $ for $k \geq 2,\ $ at the corresponding values
of $q$ which depend on both $n_{1}$ and $k$;
(ii) the position and momentum operators $X$ and $P$ {\em commute
on the state} $|n_1\rangle$ if $q$ is fixed as $q={\frac{n_1}{n_1+1}}$,
that implies unusual uncertainty relation;
(iii) two commuting copies of the creation, annihilation,
and number operators of this $q$-oscillator generate  
the corresponding $q$-deformation of the {\em non-simple} Lie
algebra $su(2)\oplus u(1)$ whose nontrivial $q$-deformed
commutation relation is: $[J_+, J_-]=2 J_0~q^{2J_3-1}$ where
$J_0\equiv \frac12 (N_1-N_2)$ and $J_3\equiv\frac12 (N_1+N_2)$.
\keywords{$q$-deformed oscillator; degeneracy of energy levels;
$q$-deformed algebra.}
\end{abstract}

\ccode{PACS Nos.: 02.20.Uw, 03.65.G, 03.65.Fd, 05.30.Pr}

\section{Introduction}

Since the famous                                       papers\cite{Bied,Macf},
the diverse aspects and applications of
Biedenharn-Macfarlane (BM) $q$-deformed
oscillators have become a very popular subject.
This same can be said about the somewhat earlier introduced
and slightly more simple Arik-Coon (AC)                        version\cite{AC}
of $q$-deformed oscillator.
The principal difference between these two versions is that
unlike the latter, the former one admits not only real,
but also the phase-like complex values of the deformation
parameter $q$, and this has important 
consequences.
For instance, a possibility of 'accidental' degeneracies
does appear in case of $q$ being a 
root of unity, popular values  
for the BM type $q$-oscillator.
Namely, for $q=\exp(\frac{{\rm i}\pi} {2n+2})$ the following
two neighboring energy levels coincide: $E_{n+1}=E_{n}$.
This and other coincidences lead to a kind of periodicity and
naturally make the corresponding
phase space both discrete and                            finite\cite{bonats}.

The question arises: {\it  is it possible that analogous property of
'accidental' degeneracy may occur at real value(s)} of the
$q$-parameter for, maybe, some other type of $q$-deformed
oscillator?  Say, is it possible that fixation of the deformation
parameter by some real value(s), e.g. $q=\sqrt{0.98}$ or
$q=1.02^{-1/2}$, does provide degeneracy of respective energy
levels?
The goal of this note is    
to present an analysis of such an exotic $q$-oscillator
and to show it possesses the already mentioned
degeneracy properties, and also a couple of other
unusual properties.
This type of $q$-oscillator has first appeared in    Ref.~\refcite{Odaka,Jagan}
where it was called the 'Tamm-Dancoff cutoff'
deformed oscillator. Representation-theoretic,
unified view, and generalized ($f$-oscillator)
aspects of the $q$-oscillator algebra
were studied in                             Ref.~\refcite{Ques1,Melj,Mizrahi}.
Our interest to this $q$-oscillator has emerged
in the course of our recent study of possible           application\cite{SIGMA}
of the (set of) two-parameter $q,\!p$-deformed
oscillators (those include the already mentioned
BM and AC $q$-oscillators as the two particular
$p=q^{-1}$ and $p=1$ limiting cases).  Namely,
the $q,\!p$-deformed oscillators have been utilized
in the framework of the model of two-parameter deformed $q,\!p$-Bose
gas model, for which the explicit formulas for
(intercepts of) general $n$-particle momentum correlation
functions have been                                       obtained\cite{AdGa}.
Note that these results generalize previously known
formulas for two-particle correlations in the AC and BM
versions of $q$-Bose gas model.

\section{Setup on the $q$-oscillator}

The {$q$-deformed oscillator} of our main interest here
is defined by  the relations:
\begin{equation}
 a a^\dagger - q a^\dagger a = q^{N} \ ,
\end{equation}
\begin{equation}
[{ N},a]=- a    \ ,     \qquad
[{ N},a^\dagger]= a^\dagger   \ .   
\end{equation}
To state the crucially important relation between
the number operator $N$, on one side, and the operators
$a^\dagger a$ or $a a^\dagger$, on the other, we appeal
to the well-known generalized two-parameter                 family\cite{Chakra}
of $p,q$-deformed oscillators defined by the relations
\begin{equation}
 A A^\dagger - q\ A^\dagger A = p^{\cal N}  \ ,  \qquad\quad \quad
 A A^\dagger - p\ A^\dagger A = q^{\cal N} \ ,
\end{equation}
supplemented with two relations involving ${\cal N}$
(completely analogous to (2)).

From the pair of relations (3) invariant under $q\!\leftrightarrow\!p$,
the formulas follow:
\begin{equation}
A^\dagger A = [\![{\cal N}]\!]_{qp} \ ,
                   \quad
                   \quad
A A^\dagger = [\![{\cal N}+1]\!]_{qp} \ ,
                   \quad
                   \quad
[\![X]\!]_{q,p} \equiv \frac{ q^{X}-p^{X} }{ q-p } \ .
\end{equation}
At $p\!=\!1$ this two-parameter system reduces
to the                                                        AC-type\cite{AC}
of $q$-bosons, while putting $p=q^{-1}$ yields
the other already mentioned                               case\cite{Bied,Macf}
of $q$-oscillators.

At $p=q$, each of the relations in (3) turns into
the single relation (1) what precludes connecting
of $A^\dagger A$ or $A A^\dagger $ immediately
with ${\cal N}$.
Besides, the $q,\!p$-bracket in (4) becomes undefined if $p=q$,
for $X$ being either an operator or a number.
Only for non-negative integer $k$ it makes sense, since
\begin{equation}
[\![k]\!]_{q,p} = \frac{ q^{k}-p^{k} }{ q-p } =
\sum_{r=0}^{k-1} q^{k-1-r}p^r =
q^{k-1}\sum_{r=0}^{k-1}q^{-r} p^r \ \ \ \
\stackrel{p\to q}{\Longrightarrow} \ \ \ \ k q^{k-1}\ .
\end{equation}
Then, let us take the analogue of the last part of relation (5)
as the (at first, formal) definition of the new $q$-bracket
in all cases, both for operators and for numbers:
\begin{equation}
\{X\}_q \equiv X q^{X-1} \ .
\end{equation}
To find more strict justification for the definition (6), let us
consider the one-parameter subfamily of $q,p$-oscillators (3)-(4).
Namely, let $p=q^m$ with real $m$:
\begin{equation}
A A^\dagger - q\ A^\dagger A = q^{m \cal N}    \qquad\quad \quad
 A A^\dagger - q^m \ A^\dagger A = q^{\cal N} \ ,
\end{equation}
\[
A^\dagger A = [\![{\cal N}]\!]_{q,q^m} \ ,
                   \quad
                   \quad
                   \quad
A A^\dagger = [\![{\cal N}+1]\!]_{q,q^m} \ .
\]
In the limit of $m\to 1$ this reduces to the q-oscillator (1),
so we consider    
\begin{equation}
{\lim}_{{m}\to 1} [\![X]\!]_{q,q^m} \equiv
{\lim}_{{m}\to 1} \frac{ q^{X}-q^{m X} }{ q-q^m } \ .
\end{equation}
It is useful to merely put $\widetilde{q}\equiv q^{m-1}$,
so that $\widetilde{q}\ \stackrel{m\to 1}{\Longrightarrow} \ 1$. Then,
\begin{equation}
[\![X]\!]_{q,q^m} =   
q^{X-1}\ \frac{ \widetilde{q}^{ X} - 1}{\widetilde{q} - 1} \ .
\end{equation}
So we define
\begin{equation}
\{X\}_q \equiv
q^{X-1} \cdot {\lim}_{\widetilde{q}\to 1}
\left( \frac{ \widetilde{q}^{ X} - 1}{\widetilde{q} - 1}
\right) =  q^{X-1} \ X \
\end{equation}
getting the same as in (6) (we have used the L'Hospital rule
in order to resolve the uncertainty).

Taking into account the formulas
\begin{equation}
a^\dagger a =  \{N\}_q \equiv N q^{N-1} \ ,
\qquad \quad
a a^\dagger = \{N+1\}_q \equiv (N+1) q^{N} \ ,
\end{equation}
we verify that the defining relation (1) is satisfied.

Note that the $q$-oscillator algebra generated by
$a$, $a^\dagger$ and $N$ obeying the relations (1)-(2)
may be given a Hopf algebra structure, with
the comultiplication, antipode and counit
written out as in                                 Ref.~\refcite{Jagan,Chung}.

\section{Energy spectrum of the $q$-oscillator}

Let us study in some detail the properties of the $q$-oscillator (1).
To this end we take, in analogy with usual quantum harmonic oscillator,
the Hamiltonian of the $q$-oscillator (1) in the form
\begin{equation}
H  = \frac{\hslash\omega}{2} (a a^\dagger + a^\dagger a) \
\end{equation}
and from now on put ${\hslash\omega}=1$ for simplicity.

We adopt the $q$-analogue of the Fock space with the
vacuum state $|0 \rangle$. Then,
\begin{equation}
a |0 \rangle = 0 \ , \qquad \quad
|n \rangle =
\frac{(a^\dagger )^n}{\sqrt{\{n \}_q!} } |0 \rangle  \ , \qquad \quad
N |n \rangle = n~|n \rangle \ ,
\end{equation}
where $\{n \}_q!=\{n \}_q\{n-1 \}_q ... \{2 \}_q\{1 \}_q$,
curly brackets are defined in (11), and
the creation/annihilation operators act by the formulas  
\begin{equation}
a \ |n \rangle = \sqrt{n q^{n-1}} \ |n-1 \rangle\ ,  \qquad  \quad
a^\dagger \ |n \rangle = \sqrt{(n+1) q^{n}} \ |n+1 \rangle \  .
\end{equation}
For any real nonnegative $q$, the operators $a$ and $a^\dagger$ are
indeed conjugates of each other.

From (12)-(14), the spectrum $H|n\rangle = E_n |n\rangle$ of
the Hamiltonian reads:
\begin{equation}
E_n =  \frac12 \Bigl( (n+1) q^{n}  +  n q^{n-1} \Bigr)
= \frac12 \ q^{n} \Bigl( 1  +  n (1+ q^{-1}) \Bigr) \ .
\end{equation}
At $q\to 1$ we recover the familiar result $E_n = n + \frac12$
as it should.
Moreover, at $n=0$ we have $E_0 = \frac12$ whatever is the value of $q$.

For any $q\ne 1$, the {\em spectrum is not uniformly spaced} (not equidistant).
It is easy to see that if $q> 1$ the spacing $E_{n+1}-E_{n}$ gradually
increases with growing $n$, so that $E_{n} \to \infty$ as $n \to \infty$.
Different, and more interesting situation emerging at $0<q<1$ is
the subject of our study in the next sections.

\vspace{1mm}
\section{ Various 'accidental' degeneracies $E_{n_1}\!=E_{n_2}$ of energy levels }

From now on we deal with the values of $q$ such that
\begin{equation}
0 <  q < 1  \ .
\end{equation}
We have to stress that the special feature of the $q$-oscillator
defined by the relations (1)-(2), with any fixed value of the
$q$-deformation parameter from (16), consists in the asymptotical
tending, as $n\to\infty$, of the $n$-th level energy to zero :
$E_n\to 0^+$.  For that reason the $q$-oscillator (1)-(2) has been
named the "Tamm-Dankoff cutoff" oscillator\cite{Odaka,Jagan}. In
this letter we use this same term (or "TD" in short).

{\bf Remark~1.} If one imposes the                     requirement\cite{Chung}
that for all the energy values the
inequality $\Delta E (n)\equiv E_{n+1}-E_n > 0$
{\em must hold}, then the spectrum is truncated and
becomes finite: the set of energy levels consists
of $E_n$ with $n=0, 1, ..., \lfloor\frac{1+q^2}{1-q^2}\rfloor$
(here $\lfloor x \rfloor$ denotes integral part of $x$).
We however do not require this in our paper, and consider
the whole (infinite) spectrum.

The energy spectrum given by the expression (15) manifests
some sorts of degeneracies, with strong dependence
on the particular fixed value of $q$.
Let us consider a number of different cases.

\vspace{1mm}
\subsection{Degeneracy of nearest-neighbor levels: $E_m = E_{m+1}$}
\vspace{1mm}

Let the $q$ be fixed as $q=\sqrt{\frac13}$.
\noindent
Then it turns out that  
\[
E_0=\frac12 <  E_1 = E_2 = \frac12 + \frac{1}{\sqrt{3}}\,,
\]
\[
E_2 > E_3=\frac12 +\frac{2}{3\sqrt{3}} \ >
E_4=\frac{5}{18} +\frac{2}{3\sqrt{3}} > \frac12\ , \quad
\]
and for the rest (infinite set) of levels we find
\[
\frac12 > E_5=\frac{1}{2}+\frac{\sqrt{3}-2}{9} > E_6 > \ldots >
E_{m} > E_{m+1} > \ldots >0 \ , \quad \qquad  m=7, 8, \ldots \ .
\]
This degeneracy $E_1 = E_2$ is a particular case of the next more general property.

{\bf Proposition 1.} Let the parameter $q$ be fixed as $q=\sqrt{\frac{m}{m+2}}$
where  $ \ m\ge 1\ $.
Then the degeneracy occurs for the following pair of the energy levels:
\begin{equation}
E_m =E_{m+1} \,.
\end{equation}
This is proved by direct check.
Figs.~1,2,3 illustrate such a degeneracy for three particular cases,
$m=1, 2, 9$.

Note that $n=0$ is excluded from the above series (17) since
the degeneracy $E_0 = E_1$ is not possible, as this would require $q=0$.
However, yet another degeneracies with the $E_0$ involved does occur, see below.
%
\begin{figure}  [hbtp]
\centering {
\includegraphics[angle=-90, width=0.65\textwidth]  {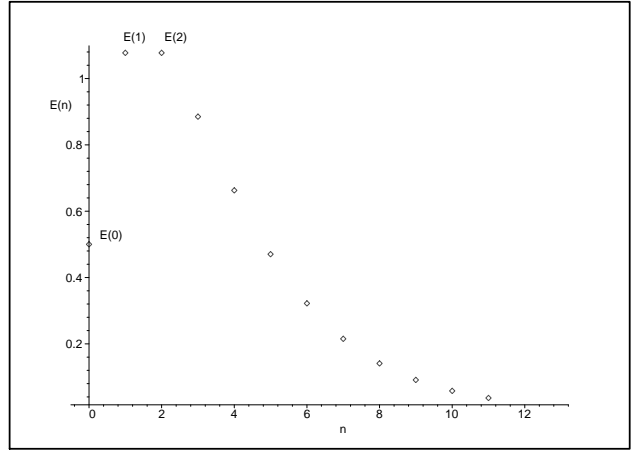}
}
\caption{ Spectrum of the $q$-oscillator (1) at fixed $q=\sqrt{1/3}$.
Observe that $E_0=\frac12$ and 
$E_1=E_2$.}
\end{figure}
\vspace{-1mm}
\vspace{-1mm}
\begin{figure}  [hbtp]
\centering {
\includegraphics[angle=-90, width=0.65\textwidth]  {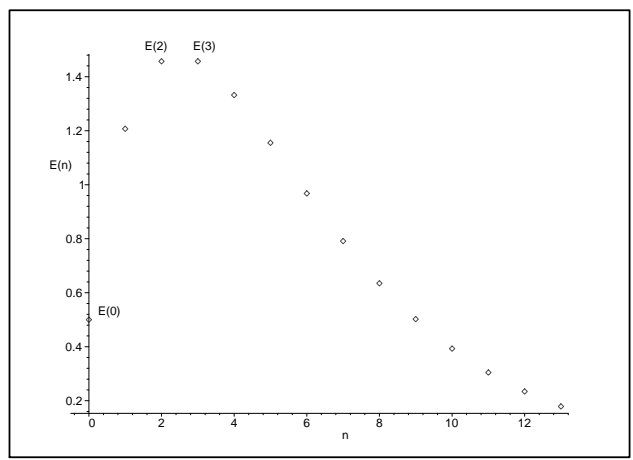}
}
\caption{ Spectrum of the $q$-oscillator (1) at fixed $q=\sqrt{2/4}$.
Observe the degeneracy 
$E_2=E_3$ .}
\end{figure}
\begin{figure}  [hbtp]
\centering {
\includegraphics[angle=-90, width=0.65\textwidth]  {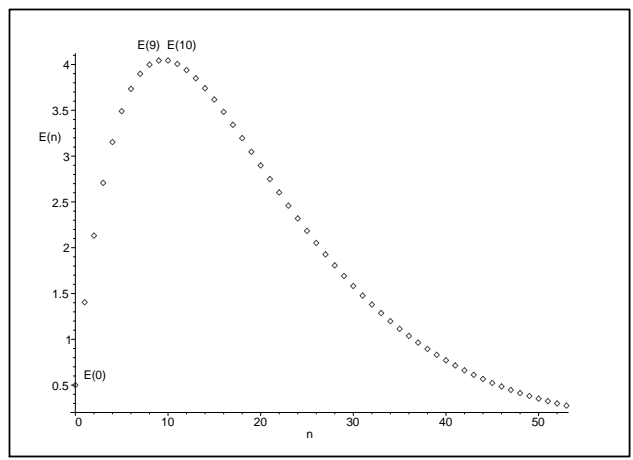}
}
\caption{ Spectrum of the $q$-oscillator (1) at fixed $q=\sqrt{9/11}$.
Observe the degeneracy 
$E_{9}=E_{10}$ .}
\end{figure}
\vspace{1mm}

\subsection{Degeneracy of next to nearest neighbors: $E_m = E_{m+2}$}

Let the $q$-parameter be fixed as $q=\frac13$.
\noindent
Then, 
\[ E_0=\frac12\ ,  \hspace{8mm}   E_1=\frac56 > \frac12\ ,
 \hspace{10mm}
E_2=E_0=\frac12\ ,   \hspace{8mm}
E_3=\frac{13}{54} < \frac12\ ,
\]
while for all other levels we find
$\ \
\frac12 > E_{m} > E_{m+1} , \ \ 
m\ge 2 \ .
$
This instance is nothing but a particular case of the next
more general situation.

{\bf Proposition 2.} Let the parameter $q$ be fixed as

\begin{equation}
q=\frac{1 + \sqrt{4 m^2 + 12 m + 1}}{2 ( m + 3)}  \qquad \qquad
                                    {\rm with}  \quad \quad   \  \ m\ge 0\ .
\end{equation}
Then among the energy levels $E_n$ the following degeneracy occurs:
\begin{equation}
E_m =E_{m+2} \ .  
\end{equation}

This statement is proved by direct verification.
The two particular cases $m=0$ and $m=5$ of the series of
degeneracies (19) are shown in Fig.~4 and Fig.~5.
\vspace{-1mm}
\begin{figure}  [hbtp]
\centering {
\includegraphics[angle=0, width=0.60\textwidth]  {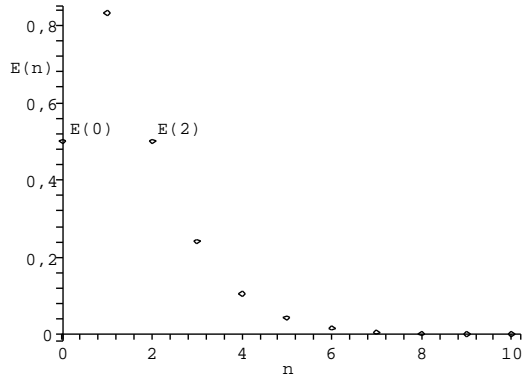}
}
\caption{ Energy levels $E_0$, $E_1$,..., $E_{10}$ of the
$q$-oscillator (1) at $q=\frac13$, see (18).
Observe $E_0=E_2$.}
\end{figure}
\vspace{-1mm}
\vspace{-1mm}
\begin{figure}  [hbtp]
\centering {
\includegraphics[angle=0, width=0.60\textwidth]  {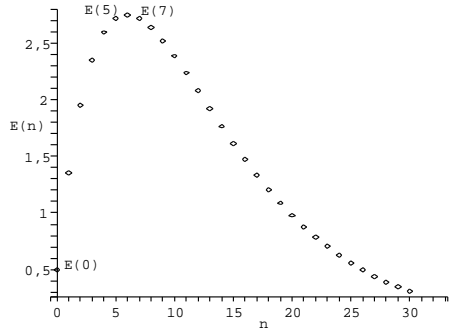}
}
\caption{ Energy levels $E_0$, ..., $E_{30}$ of the $q$-oscillator (1)
at $q=\frac{1+\sqrt{161}}{16}$, see (18).
Observe $E_5=E_7$.}
\end{figure}

\subsection{Degeneracy of the type $E_0 = E_n$}

Now let us study the occurrence of yet another type of
degeneracy,  $E_0 = E_n$.

{\bf Proposition 3.}  For any integer $ n = 2, 3, 4, ..., $
there exists the appropriate value $q_n = q(n)$ that guarantees
validity of the equality
\begin{equation}
E_0 = E_n \ .
\end{equation}
The proposition can be proved by graphical treatment.
Suppose (20) is valid. Then
\begin{equation}
(n+1) q^n + n q^{n-1} = 1\  \qquad \qquad (0 < q < 1) \
\end{equation}
or, equivalently, using for convenience $x$ instead of $n$,
\begin{equation}
 (1 + q^{-1}) x  = q^{-x} - 1\ .
\end{equation}
One can easily check the fact of intersection
(at $x$ {\em being the integers} 2, 3, 4, \ldots)
of the curves corresponding to the linear resp. exponential
functions in the l.h.s. resp. in the r.h.s. of the latter equality.
Fig.~6 and Fig.~7 serve to illustrate this fact.
\begin{figure}  [h]
\centering{
\includegraphics[angle=0, width=0.45\textwidth]  {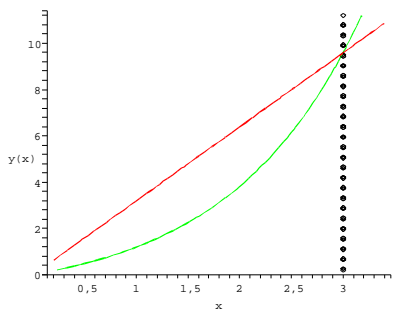}
}
\caption{ Degeneracy $E_0=E_{3}$ at $q=q_{3}\simeq 0.45541$ from
graphical solution of Eq.~(22).}
\end{figure}
\vspace{-2mm}
\vspace{-2mm}
\begin{figure}  [h]
\centering {
\includegraphics[angle=0, width=0.47\textwidth]  {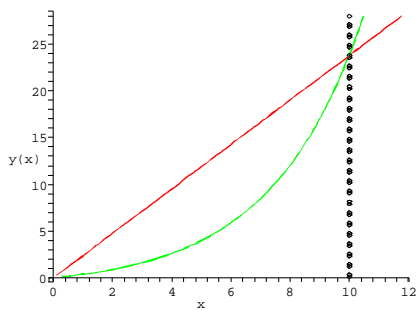}
}
\caption{ Degeneracy $E_0=E_{10}$ at $q=q_{10}\simeq 0.725405$
from graphical solution of Eq.~(22).
}
\end{figure}

\vspace{4mm}
It is instructive to have some particular values of the $q$-parameter
which provide the degeneracies of the type (20). To this end,
we rewrite Eq. (21) as the equation for $z\equiv q^{-1}$:
\[
z^m - m z - (m+1) = 0\ ,   \hspace{14mm}    m = 2, 3, 4, ... .
\]
The first three members of this series are solved exactly (in radicals),
while for $m \ge 5$ the values $q_m$ are found approximately (recall
that only those solutions $q_m$ are admitted that obey $0 < q_m < 1$).
Table 1 presents a list of sample values.
\begin{table}[h]
\tbl{Value of $q_{m}$ that gives $E_0=E_{m}$.}
{\begin{tabular}{@{}cc@{}} \toprule  
 {Value of ${m}$}  &  Value of $q_{m}$ \\ 
 \colrule
$m=2$ & \hspace{-7mm} ${q}_2= \frac13\simeq 0.333333$   \\
$m=3$ & \hspace{-5mm} ${q}_3 = \left( \sqrt[3]{2+\sqrt{3}}
                          +\sqrt[3]{2-\sqrt{3}}\ \right)^{-1}\simeq 0.45541$ \\
$m=4$ & \hspace{18mm} ${q}_4 = 3 \Bigl(1 + \sqrt[3]{64+6 \sqrt{114}}
                   + \sqrt[3]{64-6 \sqrt{114}}\ \Bigr)^{-1}\simeq 0.53156446$ \\
$m=5$  &  \hspace{-3mm}
          $q_5\simeq 0.585442$ \\
$m=6$  &   $q_6\simeq 0.6262253$ \\
$m=10$  &  $q_{10}\simeq 0.725405$   \\
$m=25$  &  $q_{25}\simeq 0.851675$   \\
$m=50$  &  $q_{50}\simeq 0.910968$   \\
$m=100$  &  $q_{100}\simeq 0.948094$   \\
$m=200$  &  $q_{200}\simeq 0.9704016$   \\
$m=400$  &  $q_{400}\simeq 0.98340363$   \\   \botrule
\end{tabular}}
\end{table}

\subsection{General two-fold 'accidental' degeneracy $E_m = E_{m+k}$}

Here we briefly discuss the most general form of two-fold
degeneracy $E_{m}=E_{m\!+\!k}$.
The equation for the $q$-parameter value $q=q(m,k)$
responsible for this fact reads
$\
(m+k+1)q^{m+k} + (m+k)q^{m+k-1}\!-\!(m+1)q^{m}- m q^{m-1} = 0\ ,
$
or
\begin{equation}
(m+k+1)~q^{k+1} + (m+k)~q^{k} - (m+1)~q - m = 0\ .
\end{equation}
It can be proved that for each pair ($m$, $m + k$) there exists
such real solution $q\!=\!q(m,k)$ of (23) that $0\!<\!q\!<\!1$.
Here we only 
comment on few cases of low $k$ values.
It is obvious that $k=1$ resp. $k=2$
correspond to the particular series of degeneracies already
considered above, in subsection 4.1 resp. subsection 4.2.
For the next two cases the equations to be solved are:
\begin{equation}
\hspace{-20mm}
\underline{k=3}  \hspace{20mm}
q^{4} + \frac{m+3}{m+4}~q^{3} - \frac{m+1}{m+4}~q - \frac{m}{m+4} = 0\ ,
\end{equation}
\begin{equation}
\hspace{-6mm}
\underline{k=4}  \hspace{16mm}
q^{4} - \frac{1}{m+5}~q^{3} + \frac{1}{m+5}~q^2 - \frac{1}{m+5}~q - \frac{m}{m+5} = 0\
\end{equation}
where for $k=4$ it was taken into account that the starting 5-th degree
equation divides by $q+1$. Note that the root $q=-1$ also exists in all
the cases of higher even $k$ in (23). The equations (24),(25) can be solved in
radicals, what yields huge expressions. And, analogously to
the above Table~1, a set of values $q=q(m,k)$
can be found numerically and as well tabulated.
Finally, let us remind that the case $E_0=E_m$, see subsec.~4.3,
is obviously covered by the most general situation (23).

\section{Fibonacci property of the TD $q$-oscillator}

Fibonacci oscillators, see                            Ref.~\refcite{Arik-fibo},
are characterized by the property
that each their energy eigenvalue $E_{n+1}$ is uniquely
determined by a linear combination of the two preceding
eigenvalues:
\begin{equation}
 E_{n+1} = \alpha\,E_{n} +  \beta\,E_{n-1}, \ \qquad  \alpha,\beta\in {\bf R}.
\end{equation}
Taking into account the explicit formula (15) 
one verifies that the TD-oscillator {\em does possess
the Fibonacci property} (26) if the coefficients are chosen as
\begin{equation}
\alpha = 2\,q  \ , \hspace{12mm}  \beta = - q^2 \ .
\end{equation}
Now it is of interest to consider the interplay
"Fibonacci + degeneracy", that is, to examine
how a specified version of the Fibonacci property manifests
itself 'locally', i.e. for those three consecutive levels
which include some degenerate pair $E_{n_1} = E_{n_2}$.

Let us first put
$q=\sqrt{\frac{m-1}{m+1}}$, $\ m>1$, which results in $E_{m-1}=E_m$.\
Then, 
\begin{equation}
E_{m+1} = q (2-q) E_m   \hspace{6mm}  {\rm or}   \hspace{6mm}
E_{m+1} = q (2-q) E_{m-1}\,, \hspace{6mm}
q=\sqrt{\frac{m-1}{m+1}}\,. \hspace{6mm}
\end{equation}
Likewise, if we
let $q=\sqrt{\frac{m}{m+2}}$ with $\ m>0$ then $E_{m+1}=E_m$.
In this case we have:
\begin{equation}
E_{m-1}=\frac{2q-1}{q^2}~E_{m} \hspace{6mm}  {\rm or}   \hspace{6mm}
E_{m-1} = \frac{2q-1}{q^2}\ E_{m+1}\ , \hspace{6mm} q=\sqrt{\frac{m}{m+2}}\ .
\hspace{6mm}
\end{equation}
Yet another situation corresponds to account of the degeneracy
of next to nearest neighbors, $E_{m+2}=E_m$, for
the $q$-parameter fixed exactly as in (18).
In this case:
\begin{equation}
E_{m+1} = \frac{1 + q^2}{2 q}\ E_{m}\  \hspace{6mm} {\rm or}
\hspace{6mm}  E_{m+2} = \frac{2 q}{1 + q^2}\ E_{m+1}
\hspace{6mm}
\end{equation}
where
$  q=\frac{1 + \sqrt{4 m^2 + 12 m + 1}}{2 ( m + 3)}\,.  $
In a similar way, other two-fold degeneracies analyzed
in Section~4 can be dealt with in conjunction with
the Fibonacci property (26).

\section{ 'Locally classical' appearance of the commutator of $X$ and $P$}

We use standard form for the position/momentum operators:
$
X\equiv \frac1{\sqrt2}(a + a^\dag), \   
P\equiv \frac{\rm i}{\sqrt2} (a^\dag - a).
$
With the account of (11), the commutator of these operators is
\[
-\frac{i}{\sqrt2}\,[ X, P ] = [ a , a^\dag ] = (N+1)q^N - Nq^{N-1} =  q^N + (q-1) N q^{N-1}\
\]
or, on the basis states,
\begin{equation}
-\frac{i}{\sqrt2}\,[ X, P ]\,|n \rangle = \bigl( q^n + (q-1) n q^{n-1}\bigr)|n \rangle\
= q^n\bigl( 1 + n(1-q^{-1})\bigr)|n \rangle\,.
\end{equation}
If we fix the deformation parameter as the rational number
$q = \frac{m}{m+1}$ ,
then it turns out that 'locally', i.e. on the individual basis
element $|n=m\rangle $, the two operators $X, P$ behave just as
the "classical" ones:
\begin{equation}
[ X, P ]\,|m \rangle = 0
 \hspace{10mm}    {\rm if}   \hspace{10mm}  q = \frac{m}{m+1}\ .
\end{equation}
Like in the cases of AC and BM type              $q$-oscillators\cite{AC,Bied}
whose $X\!-\!P$ uncertainty relations depend on both the $q$-parameter and
the state-labeling number $n$
(i.e., {\em state-dependent      uncertainty relations}\cite{Bied,M-Delgado}),
the $q$-deformed oscillator treated in the present letter also yields
the commutator $X,P$, and its associated uncertainty relation, with the
explicit $|n>$ dependence, see (31). However, unlike the AC $q$-oscillator
with {\em real} $q$, the TD type $q$-oscillator admits for each state $|m>$
the respective special value (32) of deformation parameter which makes $X$ and $P$
'classical' (commuting) on this, and only this, state.
This fact has evident consequence for (the minimum of) associated
uncertainty relation.

\section{ Quantum algebra realized with TD $q$-oscillators}

Using two identical and independent (commuting) copies of
the TD $q$-oscillator with the generators
$a_1,a_1^{\dag}\ $, $a_2,a_2^{\dag}\ $, $N_1, N_2$ obeying (1), (2), (11)
we form the generators
\[
J_+ := a_1^\dag a_2\ , \hspace{5mm} J_- := a_2^\dag a_1\ , \hspace{5mm}
J_0 : =\frac12 \bigr( N_1 - N_2 \bigr)\ ,
\hspace{3mm} J_3 : =\frac12 \bigr( N_1 + N_2 \bigr)\ ,
\]
and find their closing into the following $q$-deformed algebra
\begin{equation}
[J_0,J_{\pm}]=\pm J_{\pm} \ , \hspace{4mm} [J_+,J_-]= 2 J_0\ q^{2J_3-1}  \ ,
\hspace{4mm} [J_0,J_3]= 0 \ ,  \hspace{3mm}  [J_\pm,J_3]= 0\ ,
\end{equation}
which is a particular $q$-deformed version $U_q(su(2)\oplus u(1))$ of the
universal enveloping algebra of the {\em non-simple} algebra
$su(2)\oplus u(1)$.
This is in contrast with the well-known fact that the
$q$-analog $U_q(su(2))$ of the (simple) algebra $su(2)$
is realizable in terms of two independent modes
of the BM                                        $q$-oscillator\cite{Bied,Macf}.
The Hopf-algebra structure of the $q$-deformed
algebra (33) can be written out along the lines of        Ref.~\refcite{Chung},
so we do not reproduce it here.

Using the $q$-Fock space framework, see (13)-(14) above,
one is lead to the following rather simple yet $q$-dependent
representation formulas ($j=0,\frac12,1,\frac32,...$; $m=-j, -j+1,...,j-1,j$)
for the $q$-algebra $U_q(su(2)\oplus u(1))$:
\[
J_+ |j,m \rangle = \sqrt{(j-m)(j+m+1) q^{j-m-1} q^{j+m}}\ |j,m+1 \rangle
\]
\begin{equation}      
=q^{j-\frac12}\sqrt{(j-m)(j+m+1)}\ |j,m+1 \rangle \ ,
\end{equation}
\[
J_- |j,m \rangle = \sqrt{(j+m)(j-m+1) q^{j+m-1} q^{j-m}}\ |j,m-1 \rangle
\]
\begin{equation}    
=q^{j-\frac12}\sqrt{(j+m)(j-m+1)}\ |j,m-1 \rangle \ ,  \\
\end{equation}
\begin{equation}
J_0\ |j,m \rangle = m \ |j,m \rangle \ ,    \hspace{14mm}
J_3 \ |j,m \rangle = j  \ |j,m \rangle \ .
\end{equation}

{\bf Remark~2.} As an interesting peculiarity of this quantum algebra,
let us emphasize the following fact:  
at $q<1$, due to the multiplier $q^{j-\frac12}$ in formulas (34)-(35),
the behavior of the matrix elements of the raising/lowering operators
in the 'large spin' limit, i.e. for $j\to\infty$,
principally differs from that of $J_0$ and $J_3$. Namely, at $j\to\infty$
we have $\langle j,m |J_3|j,m \rangle \to \infty$, unlike the limit:
$\lim_{j\to\infty} \langle j,m\pm 1 |J_{\pm}|j,m \rangle = 0$.
The latter behavior is completely analogous to the (exponential) 'cutoff'
property of the TD type $q$-oscillator whose energy values satisfy:
$E_n\to 0$ when $n\to\infty$.
On the other hand such asymptotics contrasts with large $j$ behavior
of representation matrix elements of $J_+, J_-$ of the usual $su(2)$.

\section{Concluding remarks}

Our goal was to explore the rather unusual and seemingly
overlooked properties of the TD type $q$-oscillator.
In particular, we have analyzed various types of occurring
degeneracies: nearest-neighbor levels, zero level $E_0$ \&
some other level $E_m$, $m\ge 2$, etc.
The considered degeneracies are different
from the two-fold degeneracies of
oscillator-like system studied in                         Ref.~\refcite{Ques2}
where due to extra cyclic symmetry, an
infinite series of pairs of degenerate energy
levels appears. In the case of TD $q$-oscillator,
at a properly fixed value of the $q$-parameter
the degeneracy is 'accidental' (without underlying
symmetry) and involves single pair of levels.

For this, and other properties studied above,
we expect for the TD $q$-oscillator a number
of interesting physical applications.
Let us mention the already proposed                        usage\cite{SIGMA}
of TD $q$-oscillators as the base for
the related "TD $q$-Bose gas" model, the
$p\!=\!q$ one-parameter limit of $q,\!p$-Bose gas model.  As argued\cite{SIGMA},
this TD-type $q$-Bose gas model gives as well efficient
description of the observed non-Bose properties of the
two- and multi-pion (-kaon) correlations in the
experiments on relativistic heavy-ion collisions
as the other, e.g. the BM, types of $q$-Bose
gas model                                                do.\cite{AGI-1,AGI-2}

\vspace{1mm}

{\bf Acknowledgements}

One of the authors (A.M.G) thanks Prof. A.U.Klimyk for valuable
discussions and remarks.
This research was partially supported  by the Grant 10.01/015 of the
State Foundation of Fundamental Research of Ukraine.


\vspace{1mm}

\begin{thebibliography}{0}

\bibitem{Bied}
L. C. Biedenharn,
{\it J. Phys. A: Math. Gen.} {\bf 22}, L873, (1989).

\bibitem{Macf}
A. J. Macfarlane,
{\it J. Phys. A: Math. Gen.} {\bf 22}, 4581, (1989).

\bibitem{AC}
M. Arik and  D. D. Coon,
{\it J. Math. Phys.} {\bf 17}, 524 (1976).

\bibitem{bonats}
D. Bonatsos et al.,
{\it Phys. Lett. B} {\bf 331}, 150 (1994).

\bibitem{Odaka}
K. Odaka, T. Kishi and S. Kamefuchi,
{\it  J. Phys. A: Math. Gen.} {\bf 24}, L591 (1991).

\bibitem{Jagan}
S. Chaturvedi, V. Srinivasan and R. Jagannathan,
{\it Mod. Phys. Lett. A} {\bf 8}, 3727 (1993).

\bibitem{Ques1}  C. Quesne and N. Vansteenkiste,
{\it Helv. Phys. Acta} {\bf 69}, 141 (1996).

\bibitem{Melj}  S. Meljanac, M. Milekovic and S. Pallua,
{\it Phys. Lett. B} {\bf 328}, 55 (1994).

\bibitem{Mizrahi}
S. S. Mizrahi, J. P. Camargo Lima and V. V. Dodonov,
{\it J. Phys. A: Math. Gen.}  {\bf 37}, 3707 (2004).

\bibitem{SIGMA}
 A. M. Gavrilik,
{\it SIGMA} ({\it Symmetry, Integrability and Geometry: Methods and
Applications} ), {\bf 2}, paper 074, 1-12 (2006), {\tt
hep-ph/0512357}.

\bibitem{AdGa}
L. V. Adamska and A. M. Gavrilik,
{\it J. Phys. A: Math. Gen.} {\bf 37}, 4787 
                           (2004), {\tt hep-ph/0312390}.
\bibitem{Chakra}
A. Chakrabarti and R. Jagannathan,
{\it J. Phys. A: Math. Gen.}  {\bf 24}, L711 (1991).

\bibitem{Chung}
W.-S. Chung et al.,
{\it Phys. Lett. A} {\bf 183}, 363 (1993).

\bibitem{M-Delgado}
M. A. Martin-Delgado,
{\it J. Phys. A: Math. Gen.}  {\bf 24}, L1285 (1991).

\bibitem{Arik-fibo}
M. Arik et al.,
{\it Z. Phys. C} {\bf 55}, 89 (1992).

\bibitem{Ques2} C. Quesne and N. Vansteenkiste,
{\it Helv. Phys. Acta} {\bf 72}, 71 (1999), {\tt math-ph/9901016}.

\bibitem{AGI-1} D. V. Anchishkin, A. M. Gavrilik and N. Z. Iorgov,
{\it Eur. Phys. J. A} {\bf 7}, 229 (2000), {\tt nucl-th/9906034}.

\bibitem{AGI-2}
D. V. Anchishkin, A. M. Gavrilik and N. Z. Iorgov,
{\it Mod. Phys. Lett. A} {\bf 15}, 1637 (2000), {\tt hep-ph/0010019}.


\end{thebibliography}
\end{document}